\documentclass[fleqn,usenatbib,useAMS]{mnras}

\usepackage{graphicx} 
\usepackage[usenames,dvipsnames]{color}
\usepackage{latexsym}
\usepackage{amssymb}
\usepackage{amsmath}
\usepackage{soul}

\title[Halo metallicity profiles of MW-like galaxies]
{On the stellar halo metallicity profile of Milky Way-like galaxies in the Auriga simulations} 

\author[A. Monachesi et al.]{Antonela Monachesi$^{1}$\thanks{Email:antonela@mpa-garching.mpg.de}, Facundo A. G{\'o}mez$^{1}$, Robert J. J. Grand$^{2,3}$,  \newauthor Guinevere Kauffmann$^{1}$,
 Federico Marinacci$^{4}$, R{\"u}diger Pakmor$^{2}$, Volker Springel$^{2,3}$
 \newauthor and Carlos S. Frenk$^{5}$\\
  $^{1}$ Max Planck Institut f{\"u}r Astrophysik, Karl-Schwarzschild-Str. 1, D-85748, Garching, Germany\\
   $^{2}$ Heidelberger Institut f\"ur Theoretische Studien, Schloss-Wolfsbrunnenweg 35, 69118 Heidelberg, Germany\\
   $^{3}$Zentrum f\"ur Astronomie der Universitat Heidelberg, Astronomisches Recheninstitut, Monchhofstr. 12-14, 69120 Heidelberg, Germany\\
$^{4}$Department of Physics, Kavli Institute for Astrophysics and Space Research, MIT, Cambridge, MA 02139, USA\\
$^{5}$Institute for Computational Cosmology, Department of Physics, University of Durham, South Road, Durham DH1 3LE, UK}

\begin{document}

\date{}

\pagerange{\pageref{firstpage}--\pageref{lastpage}} \pubyear{}

\maketitle

\label{firstpage}

\begin{abstract}
A recent observational study of haloes of nearby Milky Way-like galaxies shows that only half (four out of eight) of the current 
sample exhibits strong negative metallicity ([Fe/H]) gradients. This is at odds with predictions from hydrodynamical simulations where  
such gradients are ubiquitous. In this Letter, we use high resolution cosmological hydrodynamical 
 simulations to study the [Fe/H] distribution of galactic haloes. We find that kinematically selected stellar haloes, including both in-situ
 and accreted particles,  have an oblate [Fe/H] distribution. 
Spherical [Fe/H] radial profiles show strong negative gradients within 100 kpc, in agreement with previous numerical results.
However, the projected median [Fe/H] profiles along the galactic disc minor axis, typically obtained in observations, are significantly flatter. The 
median [Fe/H] values at a given radius are larger for the spherical profiles than for the minor axis profiles by as much 
as 0.4 dex within the inner 50 kpc. Similar results are obtained if only the accreted stellar
component is considered indicating that the differences between spherical and minor axis profiles are not purely driven by heated disc star particles formed in situ. Our study  highlights the importance of performing careful comparisons 
between  models and observations of halo [Fe/H] distributions.

\end{abstract}


\section{Introduction}

The stellar halo metallicity distributions of massive disc galaxies contain important information about their 
formation histories. Thus, they offer a direct test of galaxy formation models. While models of stellar halos built entirely from accretion events predict mostly flat metallicity 
profiles \citep{Font06c, DeLucia_helmi08,Cooper10, Gomez12}, models which include the contribution of 
in-situ star formation predict that haloes should have strong negative metallicity gradients \citep[e.g.,][]{Font11, Tissera14}.

Hydrodynamical simulations have successfully reproduced the metallicity profile observed in M31's halo 
\citep{Font11}. However,  there seems to be some disagreement with the halo profiles observed in 
other nearby disc galaxies \citep{M13,M16}. Only half of the current sample of eight studied galaxies  (including M31 and the MW) presents strong gradients whereas the remaining 
half shows nearly flat metallicity profiles. The reason behind this discrepancy could plausibly lie in the different ways that
observed and modelled metallicity profiles are derived. Results from simulations usually present 
spherically averaged metallicity profiles \citep{Font11,Tissera13,Tissera14}. In the Milky Way (MW) and all other massive disc 
galaxies, measurements of stellar halo metallicity profiles are made only for particular lines of sight which are, most commonly, 
perpendicular to the galactic disc plane (see e.g., \citealt{Sesar11} for the MW, \citealt{Gilbert14} for 
M31, \citealt{M16} for GHOSTS galaxies), although see \citet{Ibata14} for M31.
 
 In this Letter we investigate the differences between spherically averaged and line of sight metallicity profiles
 of  stellar haloes using  high resolution cosmological hydrodynamical 
 simulations of the formation of MW-like galaxies. Thanks to their high mass resolution, these simulations 
 allow us to investigate in detail stellar halo properties of  \emph{individual simulated galaxies}, rather than 
 averaged properties from large lower resolution simulations. Our goal is to explore whether the reported tension 
 between models and observations is real or due to inconsistencies in the way they are compared. 
  
\section{Numerical Simulations}
\label{sec:sims}

We use four high resolution cosmological zoom-in simulations of MW-sized galaxies from the ``Auriga" simulation suite, 
performed using the state-of-the-art cosmological magneto-hydrodynamical code AREPO \citep{2010MNRAS.401..791S}. 
A detailed description of these simulations can 
be found in \citet[][hereafter G16]{Grand16}. Here we briefly describe their main features. 

Candidate galaxies were first selected from a parent dark matter only cosmological simulation, carried out in a 
periodic cube of side 100$h^{-1}$Mpc. A $\Lambda$CDM cosmology was adopted, with parameters $\Omega_{m}  = 0.307$, 
$\Omega_{b}  = 0.048$,  $\Omega_{\Lambda}= 0.693$,  and  Hubble  constant  $H_{0}$  = 100 $h$ km s$^{-1}$ 
Mpc$^{-1}$, $h=0.6777$ \citep{planck}.  Haloes  were selected to have masses comparable to that of 
the MW's and to be relatively isolated  at  $z=0$.  By  applying a  multi-mass  particle  
`zoom-in' technique, each halo was re-simulated at a higher resolution. 

Gas was added to the initial conditions and 
its evolution was followed by solving the Euler equations on an unstructured Voronoi mesh. The typical mass of a dark matter
particle is $\sim 3 \times 10^{5}$ M$_{\odot}$, and the baryonic mass resolution is 
$\sim 4 \times 10^4$ M$_{\odot}$. The physical gravitational softening length grows with the scale factor up to a
maximum of 369 pc, after which it is kept constant. The softening length of gas cells is scaled by the mean 
radius of the cell, with maximum physical softening of 1.85 kpc. The simulations include a comprehensive model for 
galaxy formation physics (see G16) which includes the most important baryonic processes. The model is specifically developed for the
AREPO code and was calibrated to reproduce several observables such as the stellar mass to halo mass function, the 
galaxy luminosity functions and the history of the cosmic star formation rate density. 

A summary of the properties of the galaxies analysed in this
work is presented in Table~\ref{tab:properties}. These four galaxies are a subset of the simulation suite introduced in G16
and were chosen to clearly illustrate our results. The full suite shows consistent results with those presented here and will be analysed in more detail in a follow-up work.

\begin{table}
\centering
\begin{tabular}{lrrrrrr}
\hline
Simulation & 
$M_{\rm vir}$  &
$R_{\rm vir}$  &
$M_{\rm d}$ &
$R_{\rm d}$    &
$M_{\rm b}$    &
$R_{\rm b}$ \\
\hline
Au 2  & 191.5 &  261.8 & 4.63 & 5.84 & 1.45 & 1.34 \\
Au 6  & 104.4 &  213.8 & 3.92 & 4.53 & 0.67 & 1.30 \\
Au 15 & 122.2 &  225.4 & 3.14 & 4.00 & 0.39 & 0.90 \\
Au 24 & 149.2 &  240.9 & 3.68 & 5.40 & 2.18 & 0.93 \\
\hline
\end{tabular}
\caption{Properties of the simulated galaxies at the present day. From left to right, the columns are model name (following \citealt{Grand16}), virial mass, virial radius, stellar disc mass, disc scale length, bulge stellar mass, and 
bulge effective radius. Masses are in 10$^{10}$ M$_{\odot}$ and distances in kpc.}
\label{tab:properties}
\end{table}

\section{Stellar halo definition}
\label{sec:halo}

The definition of a stellar halo is rather arbitrary and several different criteria have been used in the past 
to isolate halo from disc and bulge star particles in simulated disc galaxies. Following previous work, 
we first define the stellar halo purely based on a kinematic decomposition,
regardless of an in-situ or accreted origin for the stars. Discs are
aligned with the XY plane as in \citet{Gomez15}. For each star particle, we compute the 
circularity parameter $\epsilon = J_{z}/{J(E)}$ \citep{Abadi03}. Here $J_z$ is the angular momentum around the 
disc symmetry axis and $J(E)$ is the maximum specific angular momentum possible at the same specific binding energy, $E$. Three different 
subsets of star particles, with $\epsilon < 0.8,~0.7$ and 0.65, are selected as the spheroidal component. The contamination from particles with
disc kinematics in these subsets decreases with decreasing $\epsilon$. The first and weakest
constraint (i.e., $\epsilon < 0.8$) is equivalent to the one adopted by, e.g.,
\citet{Font11, 2012MNRAS.420.2245M, Cooper15}. The third and more restrictive constraint (i.e. $\epsilon < 0.65$) has been used
by, e.g., \citet[][]{Tissera13,Tissera14}. Following \citet{Cooper15}, particles 
from the spheroidal component that lie within 5 kpc from the galactic centre are defined to be bulge. Here we do not attempt to select halo star particles as accurately as possible. 
Instead, our goal is to replicate previously (and commonly) used selection criteria.
We also analyse a set of particles with no constraint on $\epsilon$, which is readily comparable to observations.

In addition, we analyse a second sample of star particles containing only accreted particles. This allows
us to compare with previous results based purely on the accreted component of simulated galaxies
\citep[e.g.][]{Font06c,Cooper10,Gomez12,Pillepich15}. One should bear in mind, however, that some of these
models are based on dark matter only simulations \citep[e.g.][]{Cooper10,Gomez12}. Thus, the dynamical evolution of the
baryonic component in such simulations is simplified due to, e.g., the lack of a disc gravitational potential 
\citep[see e.g.,][]{2014ApJ...783...95B}. In this work we consider a particle to be accreted if,  
at its formation time, it was bound to any subhalo other than the host. The stars that form out of gas that was stripped from 
infalling satellites and has not yet mixed with the surroundings are not included in this sample. 

We note that there is an overlap between the kinematically-selected and accreted sets of stellar particles. Since the accreted 
component of a galaxy dominates beyond 20--30 kpc \citep[e.g.][]{Abadi06, Zolotov09, Pillepich15, Cooper15}, the star particles in the outer regions ($R > 30$ kpc) are mostly the same in both samples. 
In all cases only star particles that 
at $z=0$ are gravitationally bound to the host galaxy are selected.

\section{Results}
\label{sec:res}

\begin{figure}\centering
\includegraphics[width=75mm,clip]{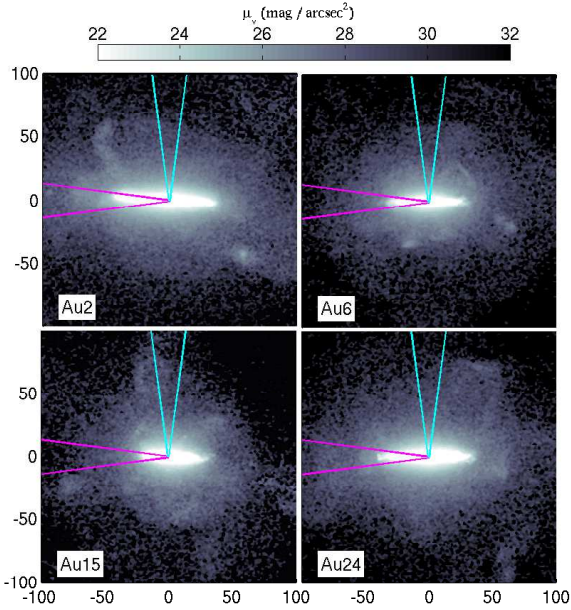}
\caption{Edge-on view of the $V-$band surface brightness profile of the four galaxies analysed in this work. Only particles that belong to the main host are considered. 
The wedges along the minor (cyan) and major (magenta) axes are used to derived the [Fe/H] profiles shown in Figs.~\ref{fig:metprofiles_circ} 
and \ref{fig:metprofiles_accre}.}
\label{fig:surfbrig}
\end{figure}

In Figure~\ref{fig:surfbrig} we show the stellar $V-$band surface brightness map of the four galaxies analysed in this work. Only star 
particles that at $z=0$ are gravitationally bound to the main galaxy are used to create the maps. 
A visual inspection reveals differences between, e.g, disc size (see Table 1) and amount of substructure and stellar halo shapes. 
The diversity in morphological properties of these MW-like simulated galaxies reflects the stochasticity inherent to the process 
of galaxy formation \citep[e.g.][]{BJ05, Cooper10}.

\begin{figure*}
\hspace{-0.2cm}
\includegraphics[width=150mm,clip]{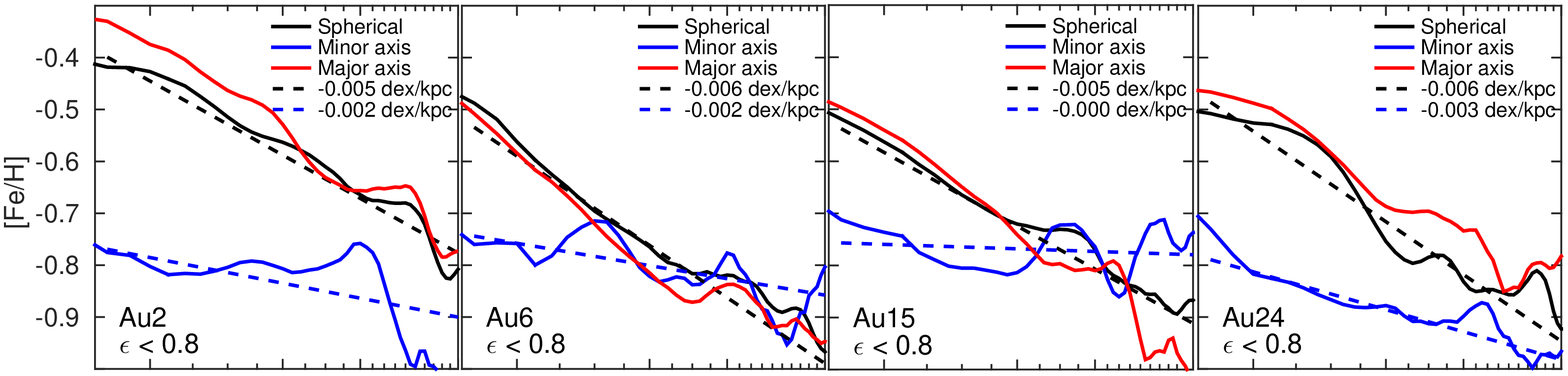}
\includegraphics[width=150mm,clip]{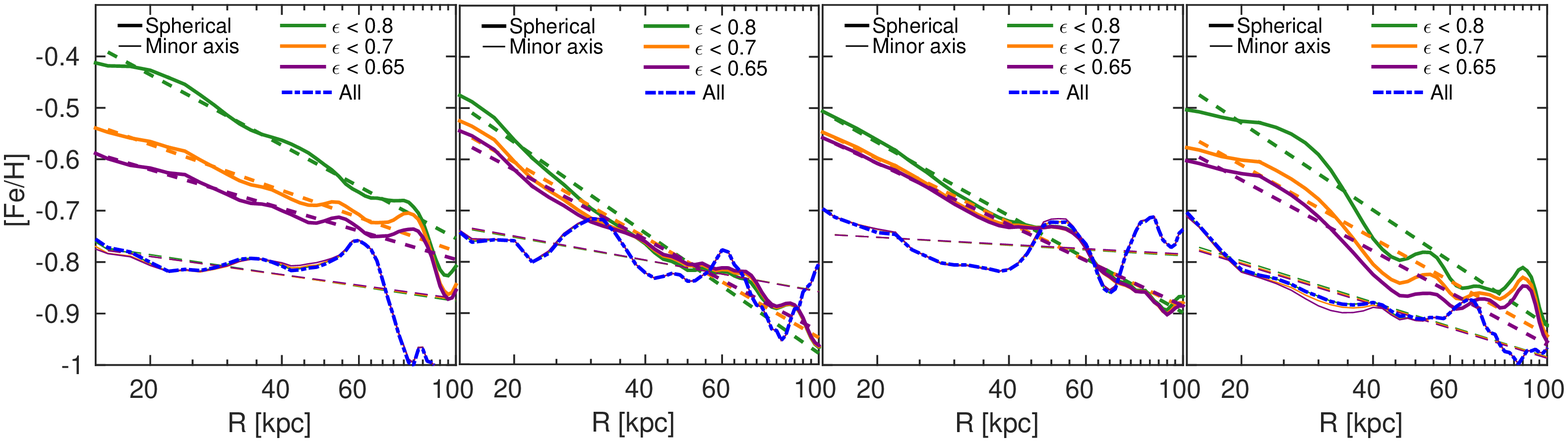}
\caption{Top: Median [Fe/H] profiles of the simulated stellar haloes between 15 and 100 kpc from the galactic centre for $\epsilon < 0.8$. 
Black lines show spherical profiles. Blue and red lines show projected profiles along the disc's minor and major axis, respectively 
(see wedges in Fig.~\ref{fig:surfbrig}).  Dashed lines are linear fits to the profiles. Their slopes in dex/kpc units are indicated in the legend. Bottom: As above for  $\epsilon <$ ~0.65 (purple), 
~0.7 (orange), 0.8 (green). The spherical and minor axis profiles are shown with thick and thin lines, respectively. 
Blue dash-dot lines show the minor axis [Fe/H] profiles using all particles, i.e. with no kinematical constraint imposed. 
We find strong differences between spherical and minor axis profiles even 
for our most restrictive selection criteria for stellar halo particles. Note that minor axis profiles are independent of $\epsilon$, so all the thin lines lie on top of one another (and to the dash-dot lines).}
\label{fig:metprofiles_circ}
\end{figure*}

Figure~\ref{fig:metprofiles_circ} shows median metallicity [Fe/H] profiles for the kinematically selected stellar halo stars. 
The profiles are shown between 15 and 100 kpc from the galactic centre. These are the regions 
generally targeted by observations of external stellar haloes, given the difficulty of isolating halo stars in the inner regions 
of a disc galaxy. The black lines show the overall median 
[Fe/H] profile, computed in spherical shells around the galactic centre. We refer to this as the spherical profile. The 
blue and red solid lines show median [Fe/H] profiles along the minor and major axes respectively. These are 
computed in $15^{\circ}$ projected wedges, as illustrated in Figure~\ref{fig:surfbrig}. To increase the number of particles 
and smooth out sudden variations due to the presence of substructure, we include on both axes particles located 
within the diametrically opposed wedge. As already stated, observed stellar halo [Fe/H] profiles  are typically obtained 
along galaxy minor axes \citep[e.g.][]{Sesar11, Gilbert14, M16, Peacock15}. 
The spherically averaged [Fe/H] profiles generally presented in theoretical works \citep[e.g.,][]{Font11,Tissera13, Tissera14} cannot 
currently be measured either for external galaxies or for the MW.  

The top panels of Fig.~\ref{fig:metprofiles_circ} show the results obtained with $\epsilon < 0.8$. These expose very 
significant differences between the spherical (black solid lines) and minor axis (blue solid lines)  $[\rm{Fe/H}]$ profiles. 
At least within the inner 50 kpc, the spherical profiles show larger median $[\rm{Fe/H}]$ values, which can differ by as much as 0.4 dex. 
More importantly, the radial dependence of the median $[\rm{Fe/H}]$ is also different. 
The dashed lines on each panel show linear fits to the profiles. In general, flatter $[\rm{Fe/H}]$ profiles are obtained along 
the minor axis compared to their spherical counterpart. The red solid lines show [Fe/H] profiles along the major axis. 
Clearly, in all cases, the spherical profiles are strongly dominated by the [Fe/H] profiles along the major axis, 
at least out to $\sim 60$ kpc. This indicates a flattened metallicity distribution in these stellar haloes. The difference between the major and minor 
axis profiles is mostly, although not entirely (see below), due to the contribution of in-situ heated disc star particles. These are particles formed in the disc that have been scattered into low $\epsilon$ orbits and thus classified as halo stars 
\citep[e.g.,][]{Zolotov09, Purcell10}.

The bottom panels of Fig.~\ref{fig:metprofiles_circ} show the results obtained for $\epsilon <$ 0.8 (green lines), ~0.7 (orange lines) and 
~0.65 (purple lines). This contrasts our results for increasingly conservative classifications of stellar halo particles.
Strong differences between minor and spherical [Fe/H] profiles are found even for our most restrictive selection criteria 
($\epsilon < 0.65)$. In some cases, such as galaxy Au 2, the difference between the median [Fe/H] profiles within 50 kpc becomes slightly smaller but is
still significant, $\sim 0.2$ dex. 
Large differences between the steepness of the profiles are found regardless of $\epsilon$. We also show, with blue dash-dot lines, 
the minor axis [Fe/H] profiles obtained when no $\epsilon$ selection is made. In this case, we omit the spherical profiles
as they purely reflect the metallicity distribution of the discs within the inner $\sim 40$ kpc. Interestingly, the minor axis [Fe/H] profiles are
indistinguishable, regardless of the selection criteria. This indicates that the mass fraction of in-situ heated disc stars along the minor axis
is negligible at distances larger than 15 kpc \citep[see also][]{Pillepich15}.

\begin{figure*}
\hspace{-0.2cm}
\includegraphics[width=150mm,clip]{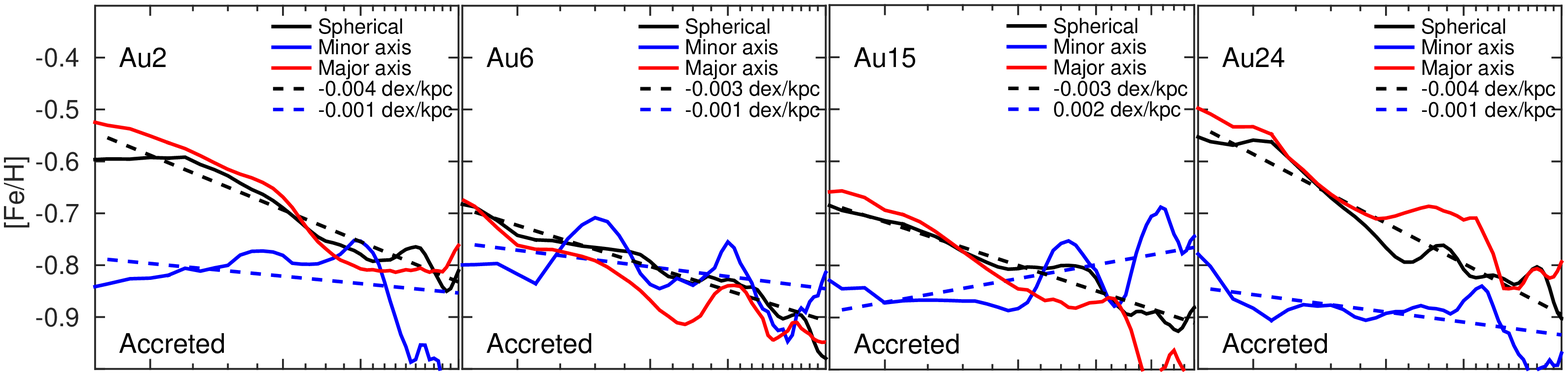}
\includegraphics[width=150mm,clip]{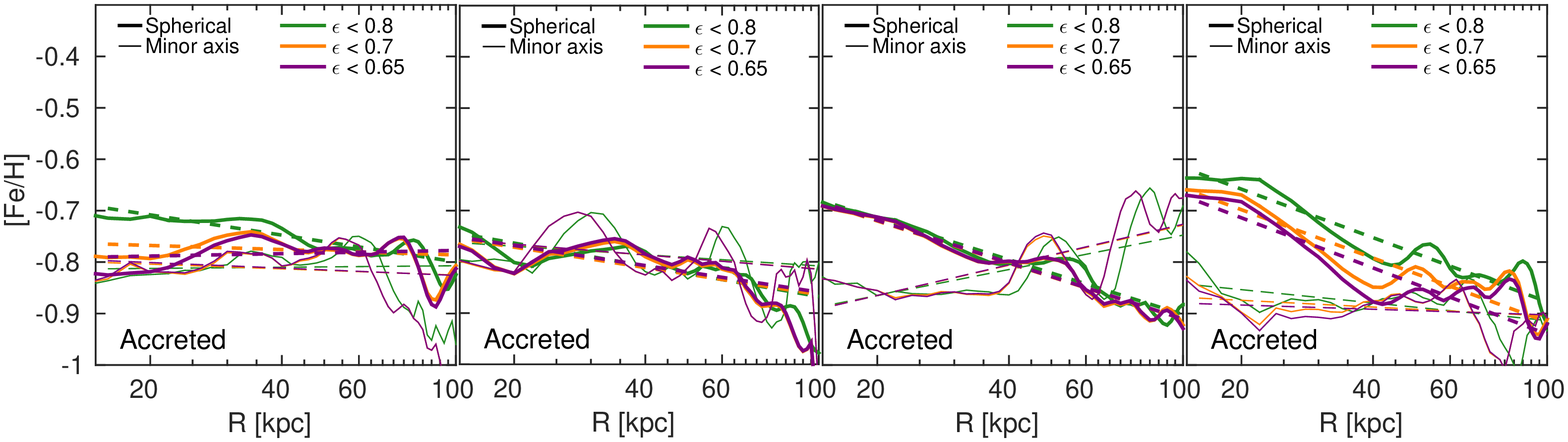}
\caption{Top: As Fig. 2 for all accreted star particles (no $\epsilon$
cut). Bottom: As Fig. 2, using three kinematic criteria to isolate accreted
stellar halo  particles. We find that for some galaxies also the accreted
component of the stellar halo shows differences in the [Fe/H] profiles when
these are derived from concentric spheres and along the minor axis.}
\label{fig:metprofiles_accre}
\end{figure*}

\begin{figure}
\centering
\includegraphics[width=75mm,clip]{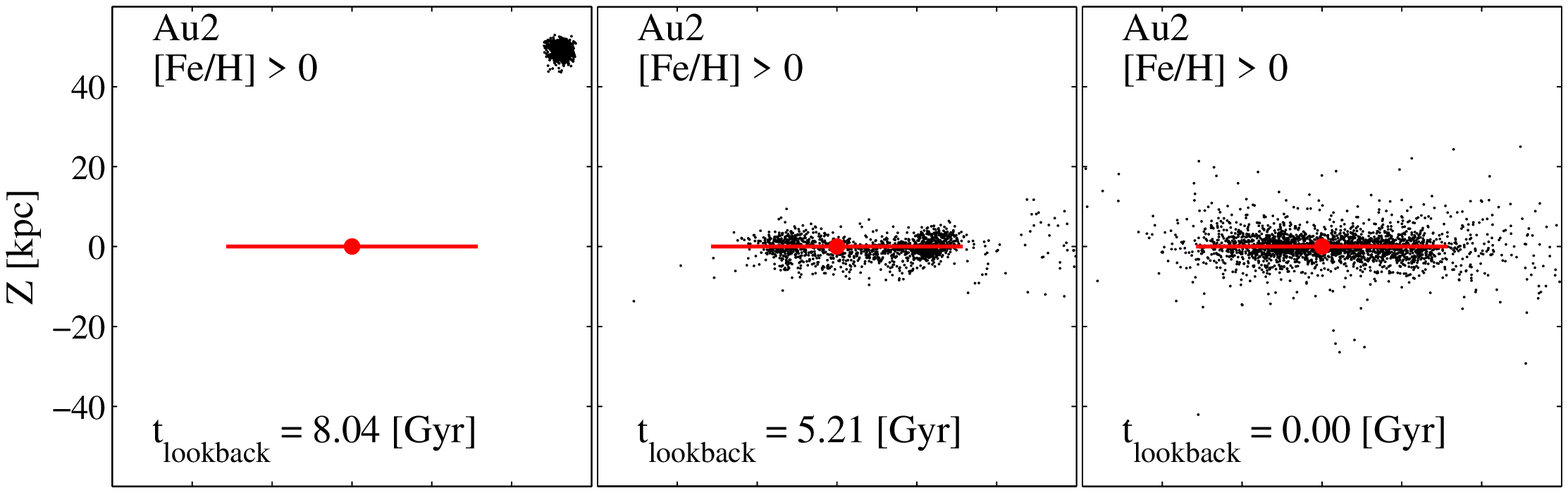}
\includegraphics[width=75mm,clip]{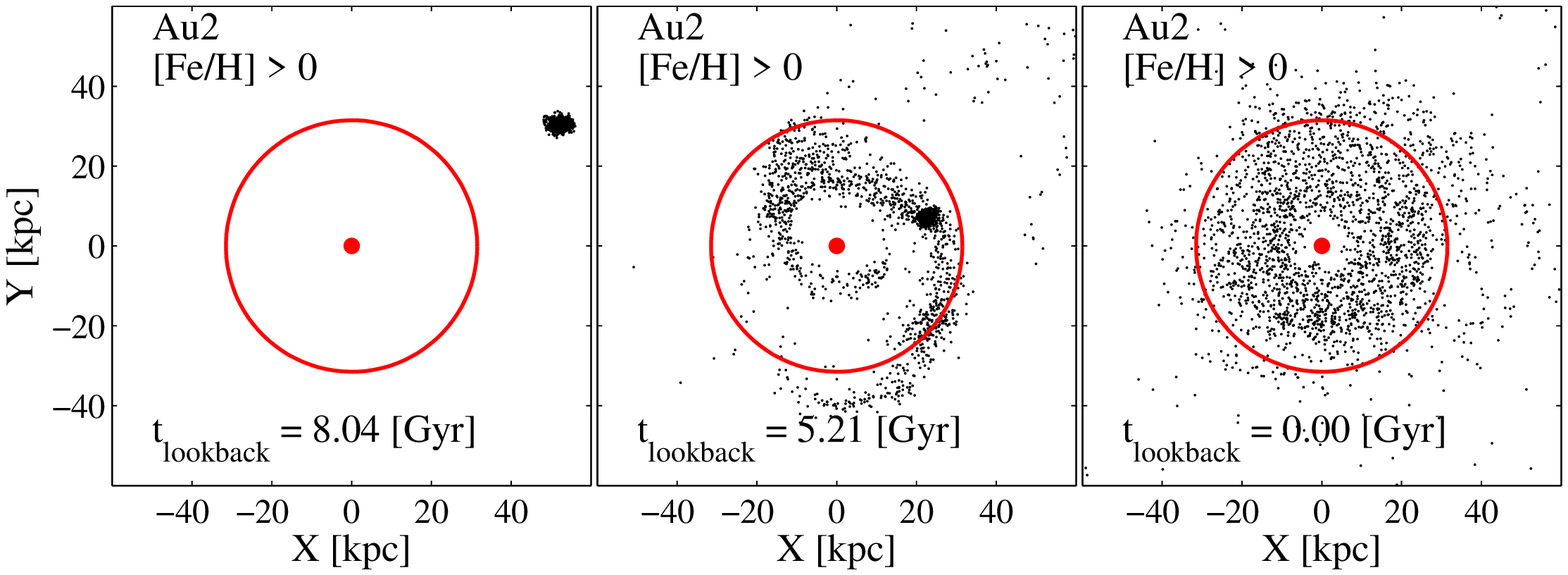}
\caption{Evolution of star particles that are more metal rich than solar of a satellite accreted into the halo of Au 2. The red lines (top panels) and circles (bottom panels) represent 8 disc scale lengths. The satellite particles end up on the disc plane very quickly, only a few Gyr after infalling into the host galaxy.}
\label{fig:sat}
\end{figure}

Figure ~\ref{fig:metprofiles_accre} shows [Fe/H] profiles taking into account only accreted particles.  
The top panels show the results obtained when no constraint in $\epsilon$ is applied. Three out of the four galaxies  show 
large differences between the spherical and minor axis [Fe/H] profiles, both in their median values and radial behaviour. Differences 
between these two profiles can be as large 0.3 dex (Au 24). This indicates that the accreted component of these simulated galaxies presents also a flattened $[\rm{Fe/H}]$ distribution. We find that as massive metal rich satellites are accreted, the host galactic disc responds 
by tilting its orientation \citep[e.g.,][]{Yurin_springel15}. As a result, in many cases the cores of massive satellites are disrupted on
a plane that is well-aligned 
with the host 
disc angular momentum \citep[see e.g.,][]{Abadi03}. Figure~\ref{fig:sat} shows one example of such accretion events. Prior to infall, at $t_{\rm lookback} \sim 8.5$ Gyr, this satellite reaches a total peak mass of $\sim 8.5 \times 10^{10}~M_{\odot}$. By $t_{\rm lookback} \sim 6$ Gyr, 
the angle between the angular momentum vector of its inner bound core and that of the disc is $< 5^{\circ}$. 

Note that, based on our kinematic selection criteria, not all the accreted particles would belong to the stellar halo. For $\epsilon < 0.8$, 
the accreted stellar mass fraction that belongs to the disc component varies between 5\% (in Au 15) and 35\%  (in Au 2).
The bottom panels of Fig.~\ref{fig:metprofiles_accre} show accreted [Fe/H] profiles using the three different 
circularities previously defined.
The differences between the spherical and minor axis [Fe/H] profiles disappear for halo Au 2 (left-most panel) in the bottom panels, indicating a strong 
orbital circularization of the inner metal rich cores of massive accreted satellites (see Fig.~\ref{fig:sat}). 
Interestingly, for haloes Au 15 and Au 24 the difference between the 
spherical and minor axis [Fe/H] profiles remains almost the same for all values of $\epsilon$. These examples show that the accreted 
component also contributes to the differences seen (both in median values and radial gradient) between the spherical and minor axis
[Fe/H] profiles. 

\section{Summary and Conclusions}
\label{sec:conclu}

While previous studies based on hydrodynamical simulations indicate that negative metallicity gradients are a 
ubiquitous feature of simulated MW-like stellar haloes \citep[e.g.,][]{Font11, Tissera13, Tissera14}, half of the current observed sample of eight nearby disc galaxies suggests flat metallicity profiles 
\citep[e.g.,][]{M16}. Motivated by this apparent discrepancy, we use four high resolution cosmological
hydrodynamical simulations of MW-like galaxies to characterise the metallicity 
profiles of their stellar haloes. Our goal is to study whether this discrepancy could be due to 
the different ways in which simulations and data are compared. In contrast to numerical studies where 
spherically averaged [Fe/H] profiles are shown, observations are obtained along a particular direction, 
ideally perpendicular to the disc plane so as to minimise disc contamination.

Following previous simulation studies, we define stellar haloes purely based on kinematic criteria. 
We find that spherical [Fe/H] profiles show large negative gradients, in agreement with previous work. 
However, significant differences are obtained when the profiles are computed in projection along the minor axis of the 
galactic disc. Not only are the median [Fe/H] values larger in the spherical profiles (up to $\sim 0.4$ dex), 
at least within the inner 50 kpc, but also the gradients are steeper in general than along the minor axis. Color profiles obtained from mock RGB stars generated as in \citet{M13} (not shown here) yield equivalent results.
We find that the spherical profiles are dominated by the halo [Fe/H] distribution along the disc's major axis. We also show
that the minor axis [Fe/H] profiles are indistinguishable for the different circularity criteria we have adopted, even when no kinematic 
selection is imposed. This indicates that a straightforward comparison between observations and models of stellar haloes is possible
along the minor axis. Furthermore, of the four examples we have analysed, one has a flat [Fe/H] profile, one has a rather steep trend and the other are intermediate. This diversity is reminiscent of the variety of profiles seen in the observational data.

Interestingly, in most cases, very similar results are obtained when only the accreted component of the stellar halo is considered, 
indicating that 
our results are not entirely driven by in-situ heated disc star particles. Note that even in dark matter only simulations, 
where baryonic effects such as a thin disc potential are not taken into account, stellar haloes generally present a flattened 
density distribution as a result of an anisotropic
distribution of satellite orbits \citep[][]{Cooper10}. Thus, those models are also expected to show differences between spherical and line-of-sight [Fe/H] profiles. 

The existence of stellar halo [Fe/H] gradients (or the lack of them) in massive disc galaxies depends strongly on how the profiles are
constructed. Such profiles depend on the adopted definition of the stellar halo, which often differs among observations 
and theoretical works. Since stellar halo [Fe/H] profiles are an important diagnostic of galaxy formation history, 
a careful and faithful comparison between observations and models is crucial in order to interpret observations correctly and constrain the 
models. This will be presented in a follow-up work.

\section*{Acknowledgments}
AM wishes to thank Eric Bell for stimulating discussions. We are grateful to Simon White for insightful comments and to Adrian Jenkins and David Campbell for the selection of the sample and making the initial conditions. We wish to thank the anonymous referee for their useful comments and suggestions. The ICs were created on the Durham DiRAC-2 Data Centric facility
supported by grants: ST/K00042X/1, ST/H008519/1, ST/K00087X/1
and ST/K003267/1. RG and VS acknowledge support through the DFG Research Centre SFB-881 
`The Milky Way System' through project A1. VS and RP acknowledge support by the European Research Council under ERC-StG grant EXAGAL-308037 and by the Klaus Tschira Foundation.

\bibliographystyle{mnras}
\bibliography{stellhalos}

\label{lastpage}

\end{document}